\documentclass[12pt]{article}

\usepackage[english]{babel}
\usepackage[utf8x]{inputenc}
\usepackage[T1]{fontenc}
\usepackage[LGR,T1]{fontenc}

\usepackage{graphicx}
\usepackage{epstopdf}
\usepackage{subcaption}
\usepackage{amsmath}
\usepackage[colorlinks=true,citecolor=black,linkcolor=black]{hyperref}
\usepackage{authblk}
\usepackage[a4paper,top=2cm,bottom=2cm,left=2cm,right=2cm]{geometry}

\usepackage[super,square, sort&compress]{natbib}
\setcitestyle{citesep={,}}

\usepackage[dvipsnames]{xcolor}

\usepackage[labelsep=period, labelfont=bf, font = {small}]{caption} 
\usepackage[section]{placeins}
\usepackage{makecell}
\usepackage{setspace} %single spaced captions 

\begin{document}

%TITLE PAGE----------------------------------------------------------------------

\title{\textbf{What Caging Force Cells Feel in 3D Hydrogels:\\ A Rheological Perspective}}

\author[1]{\textit{Giuseppe Ciccone}}
\author[1,2]{\textit{Oana Dobre}}
\author[3]{\textit{Graham M. Gibson}}
\author[1,2]{\textit{Massimo Vassalli}}
\author[1,2*]{\textit{Manuel Salmeron--Sanchez}}
\author[1*]{\textit{Manlio Tassieri}}

\affil[1]{Division of Biomedical Engineering, School of Engineering, University of Glasgow, Glasgow, G12  8LT, UK \authorcr  E-mail: Manlio.Tassieri@glasgow.ac.uk} 

\affil[2]{Centre for the Cellular Microenvironment, University of Glasgow, G12 8LT, UK \authorcr  E-mail: Manuel.Salmeron-Sanchez@glasgow.ac.uk}

\affil[3]{SUPA, School of Physics and Astronomy, University of Glasgow, Glasgow, G12 8SU, UK}

\date{}

\renewcommand{\abstractname}{\vspace{-\baselineskip}} % removing "abstract" header 

\maketitle

\newpage

%ABSTRACT-------------------------------------------------------------------------

\begin{abstract}
 \normalsize
It is established that the mechanical properties of hydrogels control the fate of (stem) cells. However, despite its importance, a one--to--one correspondence between gels' stiffness and cell behaviour is still missing from literature.
In this work, the viscoelastic properties of Poly(ethylene--glycol) (PEG)--based hydrogels -- broadly used in 3D cell cultures and whose mechanical properties can be tuned to resemble those of different biological tissues -- are investigated by means of rheological measurements performed at different length scales. When compared with literature values, the outcomes of this work reveal that conventional bulk rheology measurements may overestimate the stiffness of hydrogels by up to an order of magnitude. It is demonstrated that this apparent stiffening is caused by an induced `\textit{tensional state}' of the gel network, due to the application of a compressional normal force during measurements. Moreover, it is shown that the actual stiffness of the hydrogels is instead accurately determined by means of passive-video-particle-tracking (PVPT) microrheology measurements, which are inherently performed at cells length scales and in absence of any externally applied force.
These results underpin a methodology for measuring the linear viscoelastic properties of hydrogels that are representative of the mechanical constraints felt by cells in 3D hydrogel cultures. 
\end{abstract}

 %MAIN TEXT------------------------------------------------------------------------
 
 \newpage
 
\section*{}
Over the past two decades, a strong link has been established between the biophysical (i.e., mechanical) properties of cell--culture substrates and cell fate. \cite{engler2006matrix, leipzig2009effect, huebsch2010harnessing, rennerfeldt2013tuning, caiazzo2016defined}
Nonetheless, despite its importance, a one--to--one correspondence between gels' stiffness and cell behaviour remains still undetermined.
Hydrogels have become the most popular materials to study such relationship because of their inherent simplicity in terms of constituents and preparation, allowing a fine control of their chemical and physical properties, including to mention but a few stiffness, porosity and degradability.
Hydrogels may be broadly classified into naturally or synthetically derived materials; e.g., based on either proteins and polysaccharides (such as collagen and alginate) or PEG, respectively.
Naturally occurring polymers show a high degree of biocompatibility, with some of them being themselves constituents of the extracellular matrix (ECM); whereas, synthetic polymers offer the advantage of being rationally designed based on well defined structural units (i.e., monomers), allowing an accurate control of their biophysical properties.
Indeed, their inherent versatility has led to a conceptual shift in their design, from a simple requirement of biocompatibility to the development of bioactive materials capable to interact dynamically with their environment and to orchestrate cellular functionality such as adhesion, differentiation, proliferation and viability. \cite{huebsch2010harnessing, mann2001smooth, lee2008three, salinas2008enhancement}
A strategy to achieve this has been the incorporation of bioactive molecules such as fibronectin and laminin (LM) within the hydrogel network for mimicking the natural cellular microenvironment (where growth factors are bound to the ECM via glycosaminoglycans and other structural properties) and enhancing the hydrogel effectiveness when presented to the site of an injured tissue.\cite{salmeron2016synergistic}

Despite their successful applications in tissue engineering and regenerative medicine, it remains still difficult to draw a line between the synergistic effects of hydrogels' biochemical and biophysical properties on cell behaviour; leaving undetermined the relationships between cell fate and the stiffness of their microenvironment.
Nonetheless, it is known that cells interact mechanically with their environment in a bidirectional way; i.e., they are able either to exert forces or to perceive them from their surrounding.\cite{discher2005tissue} Moreover, while the exact molecular pathways governing this interplay remain undetermined, the exchange of forces between intra-- and extracellular environments is known to occur via focal adhesion complexes, which are large macromolecular assemblies, often of several square micrometres in area. Therefore, cells are able to `pull' their surrounding environment, probing its elasticity, via forces that are generated within the cells by means of the well--known acto--myosin interactions. Similarly, cells can `sense' forces applied to them by means of the same biotransducers, which in response activate intracellular processes.\cite{discher2005tissue, roca2017quantifying, kechagia2019integrins}

 In this regard, hydrogels have been often used to investigate cell behaviour as a function of gels' stiffness, \cite{mason2013tuning, ye2015matrix, boontheekul2007regulating, mao2016effects, wang2010role} with most of the works developed on 2D substrates, \cite{engler2006matrix} and only recently in 3D hydrogels that better resemble the natural environment of the ECM. \cite{huebsch2010harnessing, chaudhuri2016hydrogels} 
 These studies have shown a variety of cells' responses as function of gels' stiffness with variations in morphology, \cite{huebsch2010harnessing, da2014influence} spreading,  \cite{mason2013tuning, caliari2016dimensionality, sieminski2004relative} and fate. \cite{huebsch2010harnessing, pek2010effect, her2013control, goldshmid2017hydrogel}
 Moreover, it has been shown that stem cell differentiation is strongly governed by the mechanical properties of the surrounding environment more than any other biochemical factor. \cite{reilly2010intrinsic, celiz2014materials}
 Nonetheless, there is no consensus on a one--to--one correspondence between absolute values of gels' stiffness and cell behaviour, and a clear relationship is still missing in literature.
 In this study a strong experimental evidence revealing the major cause of such lack of information is provided.
 %This can be ascribed to a general malpractice in performing bulk rheology measurements of solid--like complex materials (i.e., gels).
 In particular, here the mechanical properties of a series of PEG hydrogels (functionalized with LM) have been investigated by means of both (i) bulk rheology measurements performed under different `normal force' conditions and (ii) PVPT microrheology measurements inherently performed at zero normal force.
 When compared with literature values, the outcomes of this work reveal that conventional bulk rheology measurements may overestimate hydrogels' stiffness by up to an order of magnitude.
 This is because of an induced `\textit{tensional state}' of the gel network, due to the application of a compressional normal force during the sample loading procedure. Notably, by varying the applied normal force it is actually possible to reproduce existing values of the shear elastic modulus of similar PEG--based hydrogels reported in literature. Moreover, these findings have been further corroborated by a direct comparison with PVPT microrheology measurements, which show a good agreement with bulk rheology for measurements performed at relatively low normal forces.
 
 In order to build up experimental evidences, measurements were performed on both degradable and non--degradable PEG hydrogels at PEG concentrations ranging from $3.5~\%$ to $15~\%$ w/v; which results in a range of mechanical properties that recapitulate those of a broad variety of tissues.\cite{huebsch2010harnessing, da2014influence, mason2013tuning, caliari2016dimensionality, sieminski2004relative, pek2010effect, her2013control, goldshmid2017hydrogel}
 Degradable hydrogels were obtained by incorporating the protease--cleavable peptide crosslinker GCRDVPMSMRGGDRCG (VPM); whereas, non--degradable hydrogels were obtained by incorporating the linear crosslinker PEG--dithiol (HS--PEG--SH) (see SI for details).
 PEG hydrogels have been extensively used in literature because of their hydrophilicity, biocompatibility and tunable mechanical properties. \cite{lin2009peg}
 Moreover, PEG can be easily modified with biofunctional moieties, allowing the incorporation of ECM proteins such as LMs.\cite{gilbert2010substrate, marquardt2011student, francisco2013injectable, francisco2014photocrosslinkable, ziemkiewicz2018laminin}. The latter are high--molecular weight ($400$ to $900$ kDa) heterotrimeric ECM glycoproteins (composed by $\alpha$, $\beta$ and $\gamma$ sub--units, specific to each LM isomer and arranged in a cross--like configuration) present in the basal lamina (or basement membrane) of most tissues. They are known to influence numerous cellular processes such as adhesion, differentiation, migration and survival via integrin mediated interactions; \cite{ekblom2003expression, miner2004laminin, talovic2017laminin} thus, their potential to accelerate the healing process in the presence of tissue defects and our interest in gathering a full picture of the mechanical properties of PEG--based hydrogels for future tissue engineering applications.

\begin{figure}[!hbt]
     \small
     \centering
    \includegraphics[width=\textwidth]{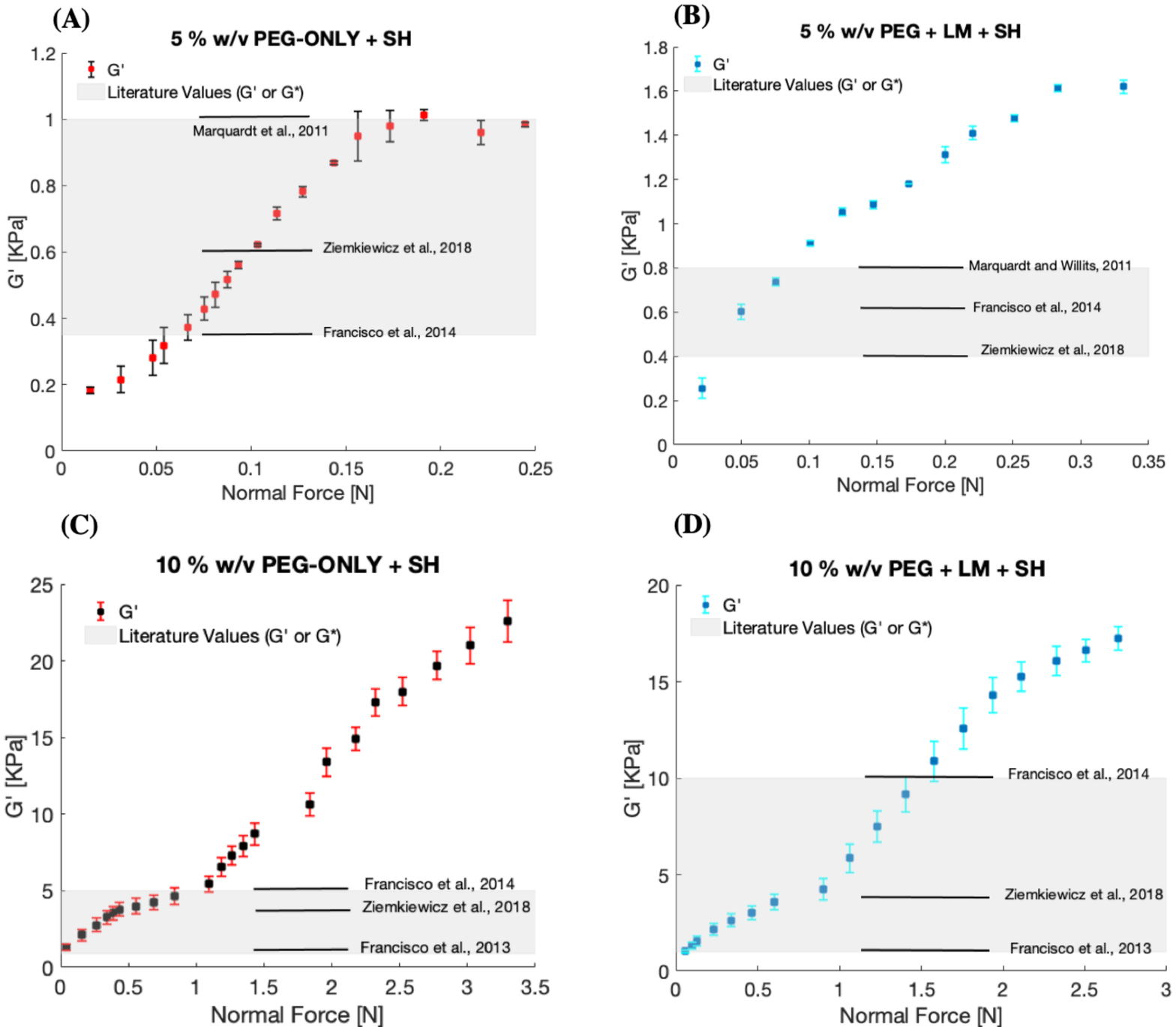}
   \caption{The shear elastic modulus ($G'$) versus the normal force for \textbf{(A)} non--degradable $5~\%$ w/v  PEG--ONLY (PEG--ONLY + SH), \textbf{(B)}  non--degradable $5~\%$ w/v PEG--LM (PEG + LM + SH), \textbf{(C)} non--degradable $10~\%$ w/v PEG--ONLY (PEG--ONLY + SH)  and \textbf{(D)} non--degradable $10~\%$ w/v PEG--LM (PEG + LM + SH) hydrogels probed via bulk rheology measurements. The results are compared with literature values reporting the mechanical properties of similar systems. Each data point has been obtained by means of strain sweep measurements performed at a frequency of $10$ rads$^{-1}$ and strain amplitudes ranging from $0.01~\%$ to $1~\%$. Error bars report $\pm 3$ standard deviations (SD). The results reveal a possible cause of the existing discrepancy between different results reported in literature, as explained in the body of the paper and here highlighted by the grey band. }
     \label{fig:Figure 1}
 \end{figure}

The rheological properties of PEG hydrogels were investigated by means of strain ($\gamma$) sweep tests, with $\gamma$ ranging from $0.01~\%$ to $1~\%$ at an angular frequency ($\omega$) of $10$ rads$^{-1}$ (see SI). A stress controlled rheometer was equipped with a parallel plate of $15$~mm in diameter. Measurements of the gels' shear elastic modulus ($G'$) were performed by gradually varying the normal force applied to the samples, starting from a minimum value of circa $0.01$~N, as shown in \textbf{Figure \ref{fig:Figure 1}}. From the latter it is clear that the mechanical properties of PEG--based hydrogels are strongly affected by the presence of a normal force, which induces a variation of $G'$ by up to an order of magnitude.
 Interestingly, when compared to literature values of similar hydrogels, \cite{marquardt2011student, francisco2013injectable, francisco2014photocrosslinkable, ziemkiewicz2018laminin} it is possible to argue that discrepancies between previous works could be simply ascribed to the presence of a normal force applied to the sample during the investigation.
 Interestingly, of the above cited works only Francisco~\textit{et al}.~\cite{francisco2014photocrosslinkable} mentioned that the measurements were performed under a compressive tare load of circa $0.01$--$0.02$~N; similar to the lower limit of applied normal force in this work.
 %Notice that, this issue would not occur  in rheological studies of viscoelastic fluids, for which a \textit{sufficiently long} time is given to the sample to fully relax after the loading procedure; this is an unachievable requirement for solid--like materials (e.g., gels), which have an infinite time of relaxation.
 
\begin{figure}[!h]
     \small
     \centering
     \includegraphics[width=\textwidth]{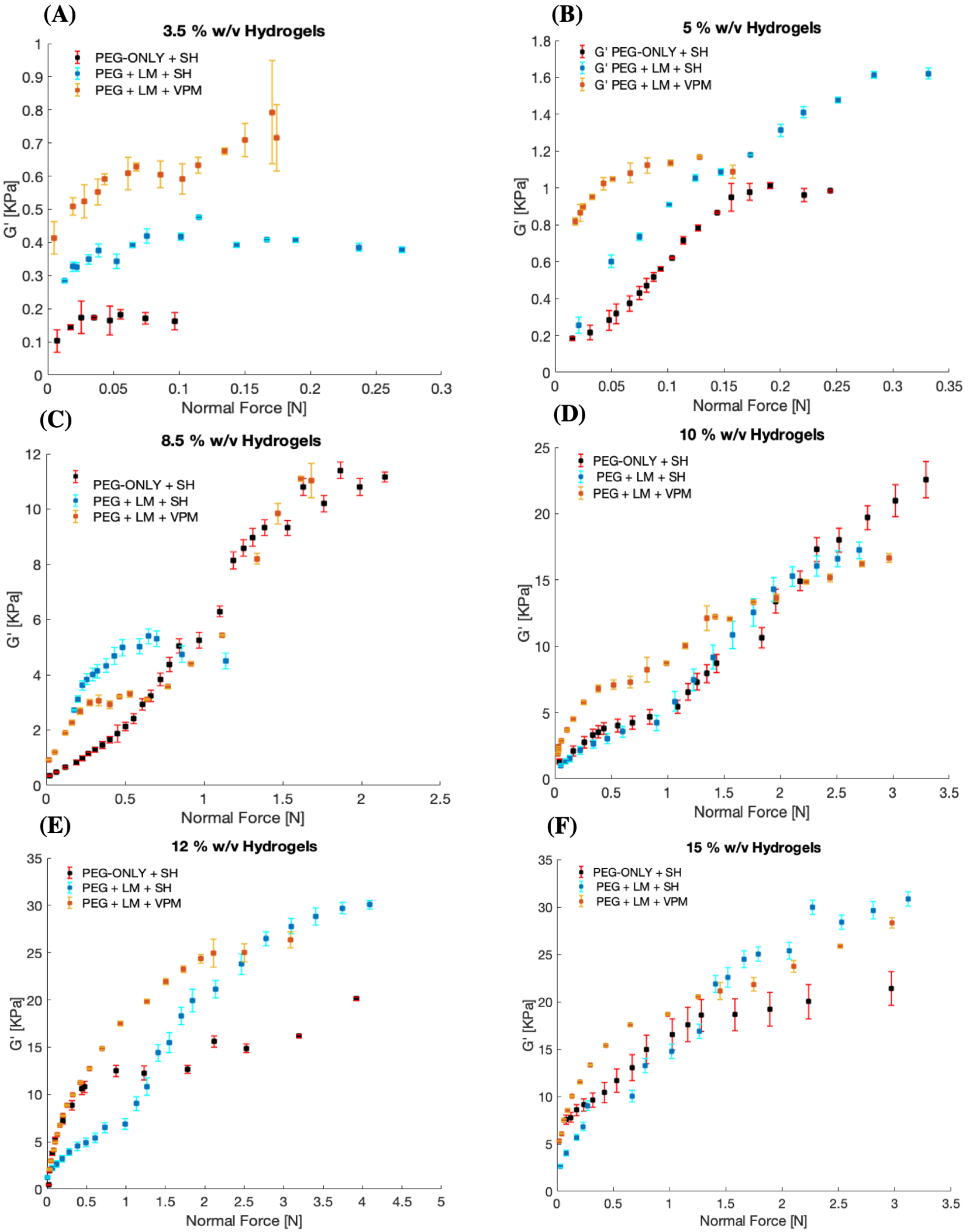}
     \caption{The shear elastic modulus ($G'$) versus the normal force measured via bulk rheology measurements for (i) PEG--ONLY non--degradable hydrogels (PEG--ONLY+SH, black), (ii) PEG--LM non--degradable hydrogels (PEG+LM+SH, blue) and (iii) PEG--LM degradable hydrogels (PEG+LM+VPM, orange) for \textbf{(A)} 3.5 $\%$ w/v, \textbf{(B)} 5 $\%$ w/v, \textbf{(C)} 8.5 $\%$ w/v, \textbf{(D)} 10 $\%$ w/v, \textbf{(E)} 12 $\%$ w/v and \textbf{(F)} 15 $\%$ w/v PEG. Each data point has been obtained by means of strain sweep measurements performed at a frequency of $10$ rads$^{-1}$ and strain amplitudes ranging from $0.01~\%$ to $1~\%$. Error bars report $\pm 3$ SD.}
     \label{fig:Figure 2}
 \end{figure}

 The increase of the shear elastic modulus as a function of the normal force is shown by all the PEG systems investigated in this work, both degradable and non--degradable systems and regardless of the functionalization with LM as reported in Figure \ref{fig:Figure 2}.
 The hypothesis is that such increase of $G'$ is due to a three--dimensional deformation of the polymeric network that results into a `\textit{strain-stiffening}' phenomenon. \cite{jaspers2014ultra} Moreover, from Figure \ref{fig:Figure 2} it is interesting to notice that such phenomenon becomes more significant as the concentration of the PEG hydrogels increases; this being inline with the increase of the number of crosslinks within the network, which promotes the stress propagation throughout the gel. \cite{ferry_1980}
 To demonstrate such hypothesis and to rule out the contribution of other possible causes that may induce a \textit{strain-stiffening} of the system, such as the loss of water because of the compression of the hydrogels, the same tests were performed on a dry, highly dense porous polymeric material (i.e., a synthetic sponge) having a diameter of $17.2$ mm and thickness $3$ mm (similar to the PEG samples) and by using two different parallel--plates having diameters of $15$ mm and $25$ mm, respectively, to investigate boundary effects.
 The results are reported in Figure S1 and they show a \textit{strain-stiffening} behaviour as for the highly hydrated gels.
 Moreover, in order to test the general validity of the hypothesis, similar tests were performed on a Polyacrylamide hydrogel  and a Polydimethylsiloxane rubber  (Figure S2), which have both shown a similar behaviour. \cite{tse2010preparation, johnston2014mechanical}
 
Further, the same methodology has been employed to shed some light on a long--standing dispute over the effects of the addition of LM to PEG hydrogels.\cite{marquardt2011student, francisco2013injectable, francisco2014photocrosslinkable, ziemkiewicz2018laminin} In particular, from Figure \ref{fig:Figure 2}, at low normal force values, LM has a significant hardening effect on softer PEG hydrogels (i.e., at PEG concentrations lower than $8.5~\%$ w/v), whereas this effect becomes less apparent for hydrogels with a higher content of PEG. Interestingly, this is not the case for PEG--LM hydrogels where the SH component has been replaced with VPM, for which a significant stiffening of the hydrogels is shown when compared to those made of PEG--only, at all the explored concentrations.

\begin{figure}[!hbt]
     %\small
     \centering
     \includegraphics[width=\textwidth]{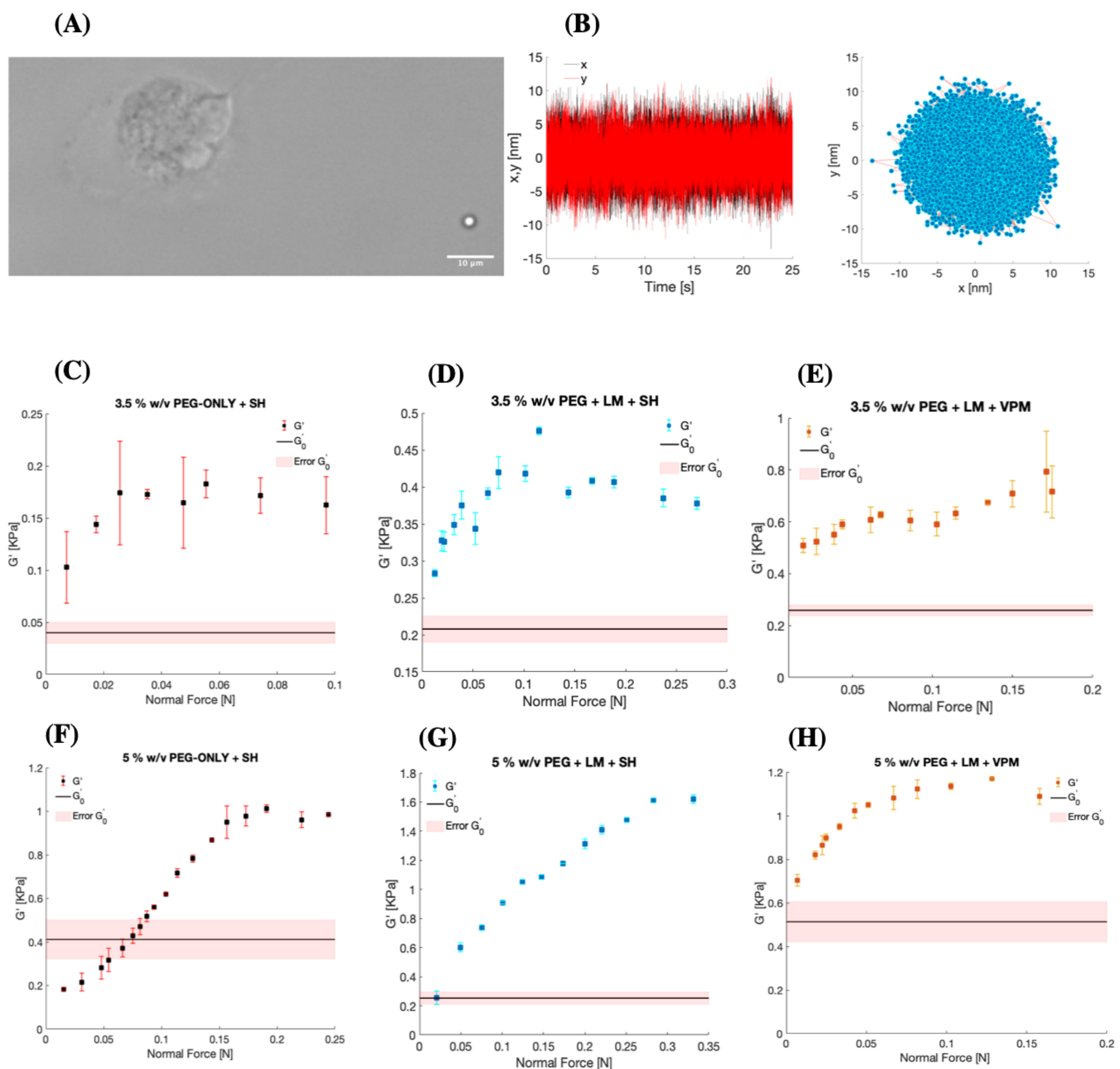}
     \caption{Comparison between bulk and microrheology measurements. \textbf{(A)} A typical image showing on the right a trapped bead within the hydrogel network (bead diameter $1.54~\mu$m) and on the left a cell. \textbf{(B)} A typical trajectory of the microsphere embedded within the hydrogel. The statistical mechanics analysis of the bead trajectory has the potential of revealing the rheological properties of the surrounding media (see SI). \textbf{(C--H)} Comparison between the shear elastic modulus, $G'$, probed via bulk rheology measurements performed at different applied normal forces, with the shear elastic modulus, $G'_0$, obtained via PVPT microrheology for the hydrogels listed in Table~\ref{tab_microrheology_results}.}
     \label{fig:Figure 3}
\end{figure}

It is now possible to draw some preliminary conclusions based on the experimental evidences reported in both Figure \ref{fig:Figure 1} and \ref{fig:Figure 2}, but also on those reported by a multitude of works aimed at finding a yet unknown bijective correspondence between gels' stiffness and cell behaviour, and from which a simple question arises: \textit{what caging force cells actually feel in 3D hydrogels?}
A possible answer to this long--standing dilemma may come from PVPT measurements, which it is reiterated are performed at the same length scales of cells and they are thermally driven, i.e., they do not require the use of any externally applied force to induce a sample deformation. In brief, the underlying principles of PVPT microrheology are based on the statistical mechanics analysis of the thermal fluctuations of micro--spheres embedded into the complex material under investigation, as shown in Figure \ref{fig:Figure 3}  (A-B) (and described in the SI). At thermal equilibrium, PVPT microrheology measurements have the potential of revealing the frequency--dependent linear mechanical properties of the surrounding media; which, in the case of solid--like materials (like gels), they can be narrowed down to a single component defined by the (almost) frequency--independent shear elastic modulus $G'$. \cite{xu1998compliance, wirtz2009particle, tassieri2019microrheology}
In Figure \ref{fig:Figure 3} (C-H) are compared the results obtained from PVPT microrheology and bulk rheology measurements, with the latter performed under different loadings of normal force.
Notice that, only relatively soft hydrogels were tested, namely those at PEG concentrations of $3.5$ and $5~\%$ w/v, as stiffer hydrogels would dampe the thermal fluctuations of the probe particle below the detector spatial resolution.
Nonetheless, the six comparisons reported in Figure \ref{fig:Figure 3} reveal the important information that cultured cells may actually feel a much lower \textit{constraining force} than generally thought; with an actual value of the gel stiffness either determined by means of PVPT microrheology measurements or estimated via bulk rheology measurements performed with relatively low applied normal forces, as summarized in Table~\ref{tab_microrheology_results}. The relatively small discrepancy between the two methods can be ascribed to the resolution of the transducer used for the detection of the normal force in bulk rheology measurements, namely $0.005$~N in this study.

To conclude, it is possible to argue that the multi--scale rheological characterization of PEG hydrogels presented in this work provides a possible justification for the \textit{yet undetermined} one--to--one correlation between the stiffness of cell culture gels and cell fate. Indeed, the experimental evidences provided in this work reveal that conventional bulk rheology measurements commonly overestimate the mechanical properties of hydrogels by up to an order of magnitude; \cite{ashworth2019peptide} therefore hindering the construction of such yearned bijective relationship. It is demonstrated that this is due to an induced \textit{tensional--state} within the gel network due to the application of a compressional normal force during sample loading; a phenomenon itself of particular importance when considering implantation of hydrogels \textit{in vivo}, as most tissues exist in a pre--stressed state.
Moreover, a direct comparison with passive microrheology measurements not only confirms the findings reported in this work, but uncovers a new strategy for the mechanical characterization of biomaterials for biomimetic culture platforms aimed at reproducing \textit{in vitro} tissue--realistic cell behaviour.

\section*{Acknowledgements}
G.C. and O.D. contributed equally to this work.
M.T. acknowledges support via EPSRC grant (EP/R035067/1 -- EP/R035563/1 -- EP/R035156/1). M.S.S. and O.D. acknowledge support via EPSRC programme grant (EP/P001114/1). We would like to thank Dr. Aleixandre Rodrigo--Navarro for his help in developing the Table of Contents graphic.

\clearpage 

%BIBLIOGRAPHY-----------------------------------------------------------------
\bibliographystyle{advancedmaterials.bst} %as required by Advanced Materials 
\newpage
\bibliography{References}

\clearpage 

%GRAPHICS-----------------------------------------------------------------------

%Figure I

 %Figure II

 %Figure III
 
%Table I

\begin{table}[!b]
    \small
    \centering
    \caption{The shear elastic modulus measured via PVPT microrheology ($G'_0$) and the minimum shear elastic modulus measured via bulk rheology ($G'_{min}$). The standard error (SE) of the mean for PVPT microrheology has been evaluated on at least four independent measurements.}
    \begin{tabular}{l l l l}
    \Xhline{4\arrayrulewidth}
    
     \textbf{Figure 3} &\textbf{Hydrogel}  &  $\text{\boldmath$G'_0 \pm$}$ \textbf{Error [Pa]}  & $\text{\boldmath$G'_{min}\pm$}$ \textbf{Error [Pa]} \\
    
    \Xhline{4\arrayrulewidth}
    
    (C) & 3.5 $\%$ PEG--ONLY+SH & 40.01 $\pm$ 10.07 & 102.7 $\pm$ 34.5 \\
    
   (D) & 3.5 $\%$ PEG+LM+SH & 207.91 $\pm$ 17.68 & 283.3 $\pm$ 4.6  \\
   
    (E) &  3.5 $\%$ PEG+LM+VPM & 258.96 $\pm$ 20.78  & 413.9 $\pm$ 49  \\
    
    (F) &  5 $\%$ PEG--ONLY+SH & 411.77 $\pm$ 89.47 & 183.3 $\pm$ 9.2 \\
     
     (G) & 5 $\%$ PEG+LM+SH & 252.88  $\pm$ 42.40 & 256 $\pm$ 45.3 \\
     
    (H) & 5 $\%$ PEG+LM+VPM & 513.32 $\pm$ 92.28 & 703.2 $\pm$ 27.5 \\
     
     \Xhline{4\arrayrulewidth}
     
    \end{tabular}
    \label{tab_microrheology_results}
\end{table}

%\newpage 
%
%%TABLE OF CONTENTS----------------------------------------------------------------------
%
%A multi--scale rheological approach reveals the true mechanical properties of Poly(ethylene-glycol) hydrogels, which are extensively used as 3D culture platforms for \textit{in vitro} and \textit{in vivo} applications. This study underpins the strategy for a quantitative evaluation of a one--to--one correlation between the linear viscoelastic properties of hydrogels and (stem) cell fate.
%
%
%\textbf{Tissue Engineering} 
%
%
%\textit{G. Ciccone, O. Dobre, G. M. Gibson, M. Vassalli, M. Salmeron--Sanchez*, and M. Tassieri*}
%
%\textbf{What Caging Force Cells Feel in 3D Hydrogels: A Rheological Perspective}
%
% \begin{figure}[hbt]
%    \centering
%    \includegraphics[width=\textwidth]{graphical_abstract.png}
%\end{figure}
%
% %suggesged dimensions: 110mm×20mm(w×h) or 55mmx50mm (w×h) 
%
% 
%\clearpage 
%
%\newpage

%SUPPORTING INFORMATION-----------------------------------------------------------------

\section*{Supporting Information}

%\subsection*{What Caging Force Cells Feel in 3D Hydrogels: A Rheological Perspective}
%
%\textit{Giuseppe Ciccone, Oana Dobre, Graham M. Gibson, Massimo Vassalli, Manuel Salmeron--Sanchez*, and Manlio Tassieri*} 
%
%\newpage 

\section*{Materials and Methods}

\textbf{\large{Hydrogel Synthesis.}} PEG--based hydrogels were synthetized via a two--step procedure. First, human recombinant LM521 (MW $= 762$ KDa, Biolamina) was PEGylated at a molar ratio of $1:25$ with Ac--PEG--NHS (MW $=2$ kDa, Laysan Bio, Inc.) to introduce functional acrylate groups. LM521 came as a phosphate--buffered saline (PBS) solution at $100~\mu$gml$^{-1}$. LM521 was  dialyzed to change the buffer to sodium bicarbonate (NaHCO3), pH $8.5$ ($30$ min at $4^{o}$C). Subsequently, to obtain the desired molar ratio, Ac--PEG--NHS ($1$mgml$^{-1}$ in NaHCO3) was added to the LM521 solution and left to mix on a steering plate for $2$h at room temperature (RT). The solution was dialyzed again to remove unreacted Ac--PEG--NHS ($30$ min at $4^{o}$C). 
In order to obtain a wide range of stiffnesses of non--degradable hydrogels (3.5, 5, 8.5, 10, 12 and 15 $\%$ w/v), different amounts of 4--Ac--PEG (MW $= 10$ kDa, Laysan Bio, Inc.), HS--PEG--SH (MW $= 2$ kDa, Creative PEGWorks) and a photoinitiator ($0.1\%$ Irgagure 2959) were mixed to the previous solution (ratio of Ac:SH$=2:1$). The solution was injected into custom Polydimethylsiloxane molds (diameter $17.2$ mm and thickness $2$ mm for bulk rheology measurements and diameter $6$ mm and thickness $2$ mm for microrheology tests) specifically designed and optimized for the mechanical characterization of the material. The gelation was carried out via photopolymerization under UV light for $5$ min at $5$ mWcm$^{-2}$ (OmniCure Series 1500, Excelitas Technologies Ltd).
The same procedure was repeated for the production of degradable hydrogels, but using 4--Ac--PEG (MW $= 10$ kDa, Laysan Bio, Inc.) and VPM (MW $= 1.7$ kDa, GenScript) (ratio of Ac:VPM$=2:1$). For PEG--ONLY gels, LM521 was replaced by PBS.
Note that, a solution of Polystyrene (PSS) particles ($1~\mu$lml$^{-1}$, particle radius $= 0.77~\mu$m, 2.5 $\mu$l per gel) was added to the gels prepared for microrheology measurements. Gels were left in PBS until complete swelling was achieved. The hydrogel synthesis process is shown schematically in Figure S3.

\textbf{\large{Bulk Rheology.}} Rheological measurements were carried out by using a stress--controlled rheometer (MCR302, Anton Paar) equipped with a parallel plate geometry, with upper plate diameter of $15$ mm, at a temperature of $23^{o}$C. Strain sweeps in the range of $0.01~\%$ to $1~\%$ and angular frequency of $10$ rads$^{-1}$ were performed to determine the viscoelastic moduli of the material. A series of strain sweep tests were performed for each sample at different normal forces. Samples' hydration was maintained by addition of PBS at the exposed sides of the specimens.
The following PEG--based gels were tested: 

\begin{itemize}
    \item non--degradable PEG--ONLY hydrogels at concentrations of $3.5, 5, 8.5, 10, 12$ and $15\%$ w/v;
    \item non--degradable PEG--LM hydrogels at concentrations of $3.5, 5, 8.5, 10, 12$ and $15\%$ w/v;
    \item degradable PEG--LM hydrogels at concentrations of $3.5, 5, 8.5, 10, 12$ and $15\%$ w/v.
\end{itemize}

In addition, bulk rheological tests were performed on a $10\%$ w/v Polyacrylamide hydrogel prepared by following the protocol described by Tse and Engler,\cite{tse2010preparation} a Polydimethylsiloxane rubber ($10:1$),\cite{johnston2014mechanical} and a dry synthetic porous material (i.e., a sponge). The sponge had diameter of $17.2$ mm and thickness of $3$ mm (i.e. dimensions similar to the hydrogel samples), and was tested by using both a $15$ mm and a $25$ mm upper parallel plates, performing strain sweeps in the range of $0.01~\%$ to $0.1~\%$ and angular frequency of $10$ rads$^{-1}$.

\vspace{0.5cm}

\textbf{\large{Microrheology.}} Measurements were performed by using a passive video particle tracking method. The apparatus comprised of an inverted microscope, with an objective lens (100X, 1.3 Numerical Aperture, oil immersion, Zeiss, Plan-Neouar) used to image the thermal fluctuations of entrapped PSS particles of $0.77~\mu$m radius within the hydrogel network. A complementary metal--oxide semiconductor camera (Dalsa Genie HM640 GigE) recorded high--speed images of a reduced field of view. Images were processed in real time at $\approx 4$ KHz using a custom made LabVIEW (National Instruments) particle tracking software. \cite{gibson2008measuring} All measurements were carried out at a constant temperature of $23^{o}$C $\pm$ $1^{o}$C.
Only the softer gels (i.e., $3.5$ and $5\%$ w/v) were tested. Stiffer gels were hindering the detection of the thermal fluctuations of the entrapped beads. The following gels were tested: 

\begin{itemize}
    \item non--degradable PEG--ONLY hydrogels at concentrations of $3.5$ and $5\%$ w/v;
    \item non--degradable PEG--LM hydrogels at concentrations of $3.5$ and $5\%$ w/v;
    \item degradable PEG--LM hydrogels at concentrations of $3.5$ and $5\%$ w/v.
\end{itemize}

\vspace{0.5cm}

\textbf{\large{Theoretical Background of Microrheology.}} When a micron--sized spherical particle is suspended into a viscoelastic fluid at thermal equilibrium, it experiences random forces leading to its Brownian motion, which is driven by the thermal fluctuations of the fluid's molecules. It has been shown that the statistical mechanic analysis of the bead trajectory $\Vec{r}(t)$ $\forall$ $t$ can be directly related to the linear viscoelastic properties of the surrounding environment by solving a generalized Langevin Equation:\cite{tassieri2019microrheology, rizzi2018microrheology}

\begin{equation}\label{eq: 1}
    m \Vec{a}(t) = \Vec{f}_R (t) - \int_{0}^{t} \zeta(t-\tau) \Vec{v}(\tau) d\tau, 
\end{equation}

\noindent where $m$ is the mass of the particle, $\Vec{a}(t)$ is its acceleration, $\Vec{v}(t)$ its velocity and $\Vec{f}_R (t)$ is the Gaussian white noise term, describing stochastic thermal forces acting on the particle. The integral term represents the viscous damping by the fluid, originating in response to the random motion of the particle due to the thermal fluctuations of the surrounding molecules, where $\zeta(t)$ is the generalized time--dependent memory function. Assuming that the Laplace transform of the bulk viscosity of the fluid $\Tilde{\eta}(s)$ is proportional to the microscopic memory function,

\begin{equation}\label{eq:2}
    \Tilde{\zeta}(s) = 6 \pi a \Tilde{\eta}(s), 
\end{equation}

\noindent where $a$ is the bead radius, one can obtain the solution to Equation \ref{eq: 1} in terms of the particle's mean square displacement (MSD): 
 
 \begin{equation}\label{eq:3}
     G^*(\omega) = s\Tilde{\eta}(s)|_{s=i\omega} = \frac{1}{6 \pi a} \Bigg[\frac{6 k_B T}{i\omega \Big \langle \Delta \hat{r^2}(\omega) \Big \rangle}+ m\omega^2 \Bigg], 
 \end{equation}
 
\noindent where $k_B$ is the Boltzmann's constant, $T$ is the absolute temperature and $\Big \langle \Delta \hat{r^2}(\omega) \Big \rangle$ is the Fourier transform of the MSD, which can be expressed as: 
 
\begin{equation}\label{eq:4}
    \langle \Delta r^2 (\tau) \rangle \equiv  \Big \langle [\Vec{r}(t+\tau) - \Vec{r}(t)]^2 \Big \rangle.
\end{equation}

\noindent The average $\langle ... \rangle$ is taken over all initial times $t$ and all particles, if more than one is observed. 
Moreover, it has been shown that there exists a link between microrheology and bulk rheology measurements. \cite{tassieri2019microrheology, rizzi2018microrheology} This is obtained by retrieving the constitutive equation between the material's shear relaxation modulus $G(t)$ and its shear compliance $J(t)$:

\begin{equation}\label{eq:6}
    \int_{0}^{t} G(\tau) J(t-\tau) d\tau = t. 
\end{equation}

\noindent where $J(t)$ is defined as the ratio of the time-dependent shear strain $\gamma (t)$ to the magnitude of a step shear stress applied at $t=0$: 

\begin{equation}\label{eq: 5}
    J(t) = \frac{\gamma (t)}{\tau_0}.
\end{equation}

Moreover, given that the shear complex modulus $G^*(\omega)$ is defined as the Fourier transform of the time derivative of $G(t)$, by taking the Fourier transform of Equation \ref{eq:6}, one obtains: 

\begin{equation}\label{eq:7}
    G^*(\omega) = i\omega \hat{G}(\omega) = \frac{1}{i\omega \hat{J}(\omega)}, 
\end{equation}

where $\hat{G}(\omega)$ and $\hat{J}(\omega)$ are the Fourier transforms of $G(t)$ and $J(t)$, respectively. \\ Finally, given that the Fourier transform is a linear operator, by equating Equations \ref{eq:3} and \ref{eq:7}, one obtains: 

\begin{equation}\label{eq:8}
    \Big \langle \Delta \hat{r^2}(\omega) \Big \rangle = \frac{K_BT}{\pi a} \hat{J}(\omega)  \hspace{0.5cm} \Longleftrightarrow  \hspace{0.5cm} \Big \langle \Delta{r^2}(\tau) \Big \rangle = \frac{K_BT}{\pi a} J(t).
\end{equation}
\\
Equation \ref{eq:8} expresses the linear relationship between the MSD of a suspended spherical particle and the macroscopic creep compliance of the surrounding fluid, therefore allowing to evaluate the fluid's complex shear modulus via Equation \ref{eq:3} without the need of a preconceived model, once an effective analytical method for performing the Fourier transform of a discrete set of experimental data is adopted, and the MSD of the random motion of the entrapped microsphere can be recorded with high spatial and temporal resolution. \cite{tassieri2019microrheology} \\ 
When a particle is embedded into a viscoelastic solid, its diffusion is highly hindered and the viscous (damping) contribution to the thermal fluctuations are significantly lower that the constraining elastic force exerted onto the particle by the gel network; therefore, the creep compliance becomes in good approximation inversely proportional to the shear elastic modulus: \cite{ferry_1980}

\begin{equation}\label{eq:9}
    J(t) \cong \frac{1}{G'(t)}. 
\end{equation}

Moreover, in the limit of purely elastic materials, it has been shown that: \cite{xu1998compliance}

\begin{equation}\label{eq:10}
    \Big \langle \Delta{r^2}(\tau) \Big \rangle \cong  \Big \langle {r^2} \Big \rangle, 
\end{equation}

Consequently, implementing the conditions expressed in Equations \ref{eq:9} and \ref{eq:10} into Equation \ref{eq:8}, one can write: 

\begin{equation}\label{eq:2.3}
     G' \cong \frac{K_BT}{\pi a \Big \langle {r^2} \Big \rangle},
\end{equation}

from which the elastic shear modulus can be easily calculated knowing the radius of the bead and the variance $\Big \langle {r^2} \Big \rangle$ of its trajectory.

\newpage

\begin{figure}[!hbt]
     \small
     \centering
     \includegraphics[width=\textwidth]{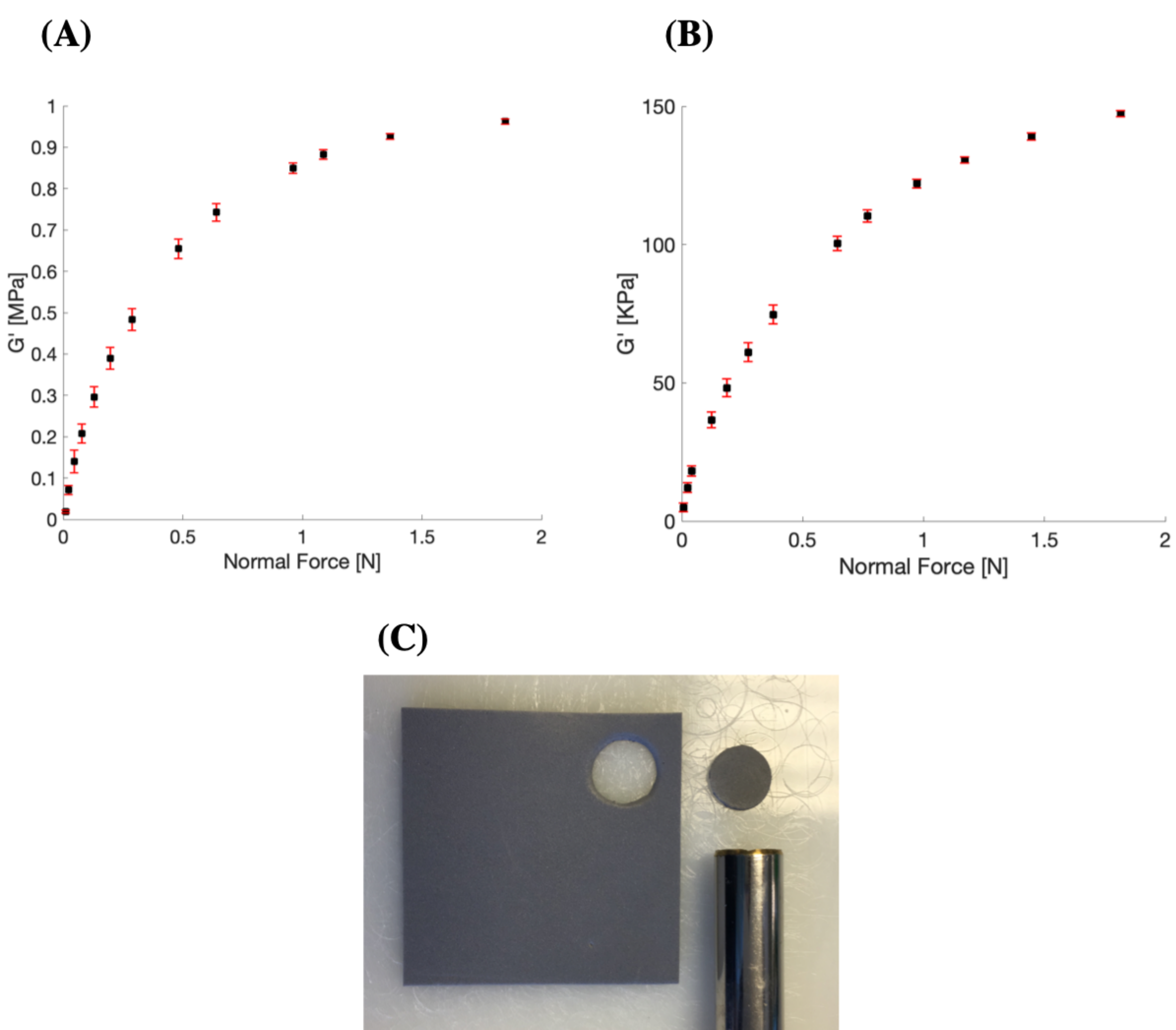}
     \caption*{\textbf{Figure S1.} The shear elastic modulus ($G'$) versus the normal force measured via bulk rheology for a dry, highly dense porous polymeric material (i.e., a synthetic sponge). \textbf{(A)} The dependency of $G'$ on the normal force when employing a parallel plate of 15 mm. Note the units of the $y$ axis, MPa. \textbf{(B)} The dependency of $G'$ on the normal force when employing a parallel plate of 25 mm, resulting in an underestimation of the elastic modulus, as expected. Note the units of the $y$ axis, KPa. Each data point has been obtained by means of strain sweep measurements performed at a frequency of $10$ rads$^{-1}$ and strain amplitudes ranging from $0.01~\%$ to $0.1~\%$. Error bars report $\pm 3$ SD. \textbf{(C)} A sample specimen. }
     \label{fig:FigureS1}
 \end{figure}

\newpage 

\begin{figure}[!hbt]
     \small
     \centering
     \includegraphics[width=\textwidth]{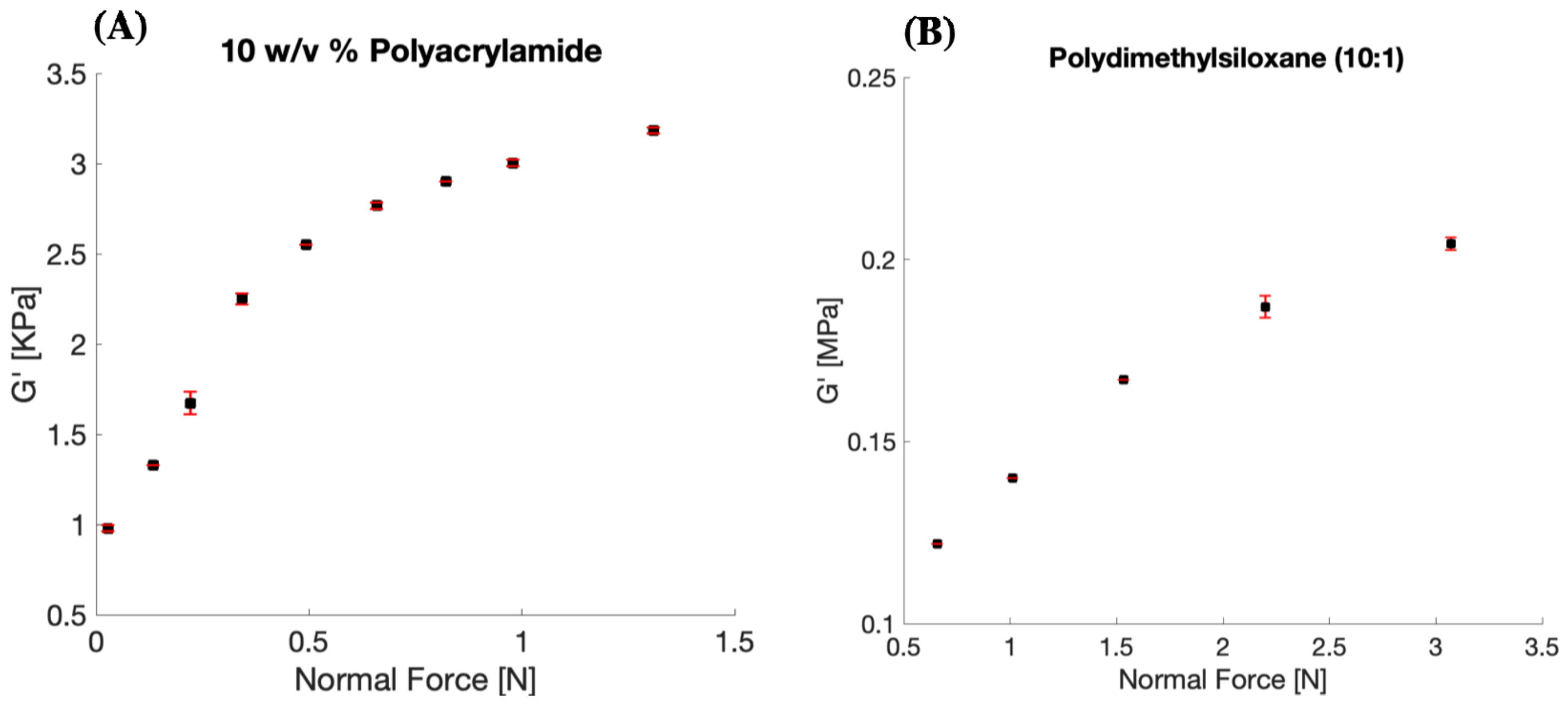}
     \caption*{\textbf{Figure S2.} The shear elastic modulus ($G'$) versus the normal force measured via bulk rheology measurements for \textbf{(A)} 10 $\%$ w/w Polyacrylamide hydrogel and \textbf{(B)} 10:1 Polydimethylsiloxane rubber. Each data point has been obtained by means of strain sweep measurements performed at a frequency of $10$ rads$^{-1}$ and strain amplitudes ranging from $0.01~\%$ to $1~\%$. Error bars report $\pm 3$ SD.}
     \label{fig:FigureS2}
 \end{figure}
 
 \begin{figure}[!hbt]
     \small
     \centering
     \includegraphics[width=\textwidth]{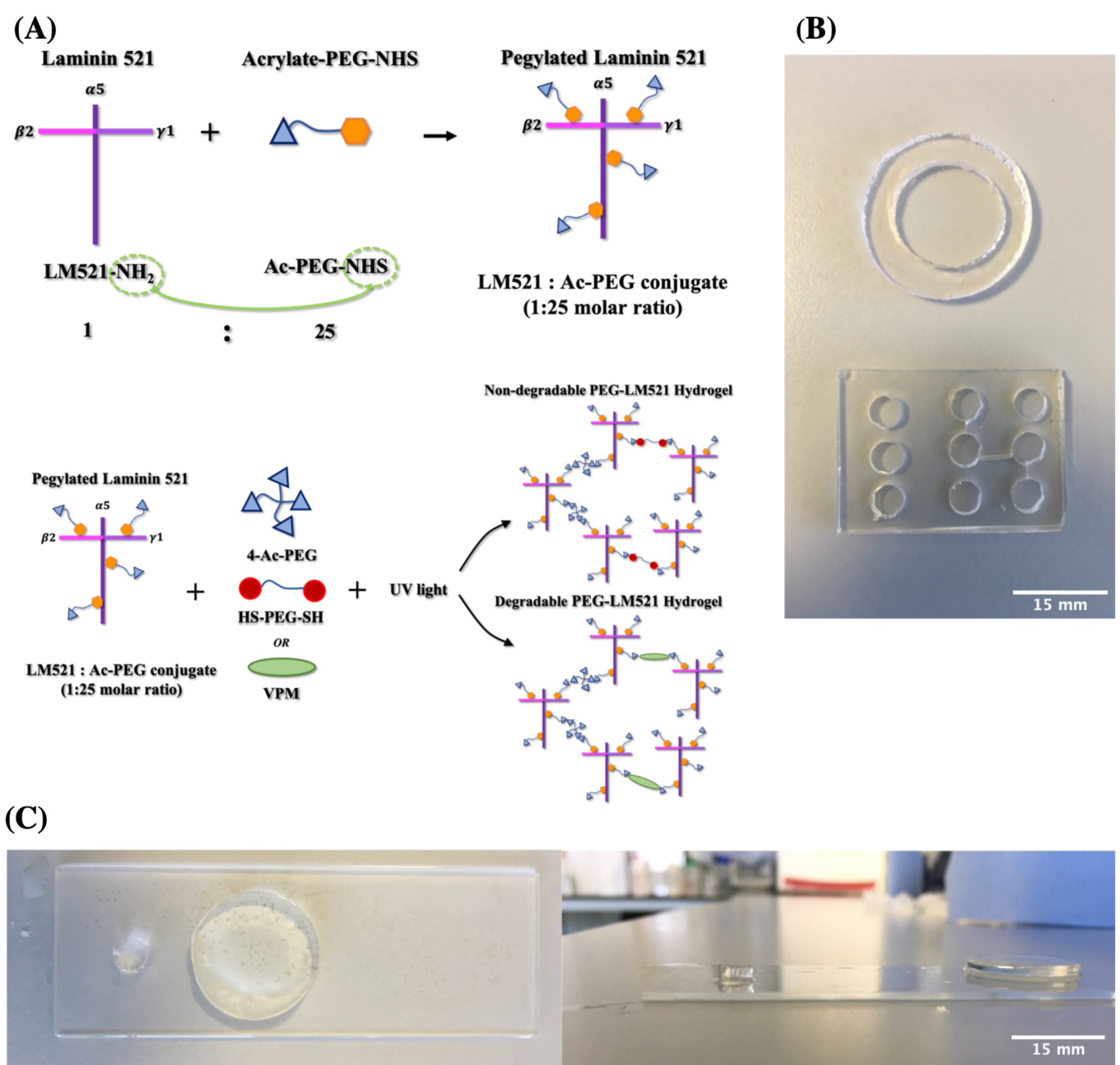}
     \caption*{\textbf{Figure S3.} PEG--based hydrogels synthesis. (\textbf{A}, top) The PEGylation reaction, where Ac (blue triangle)--PEG (blue rod)--NHS (orange hexagon) is conjugated to LM521 molecules (purple cross) in order to introduce functional acrylate groups at a molar excess of 1:25 to obtain pegylated LM521. (\textbf{A}, bottom) The formation of non--degradable or degradable hydrogels with the addition of HS--PEG--SH or VPM, respectively, and crosslinker 4--Ac--PEG via a photopolymerization reaction. \textbf{(B)} The custom--made Polydimethylsiloxane molds, top for gels intended for bulk rheology and bottom for gels intended for microrheology, respectively. \textbf{(C)} A top and side view of two sample hydrogels, respectively left and right, the largest intended for bulk rheology and the smallest for microrheology. Note the flat surface of the hydrogels.}
     \label{fig:FigureS3}
 \end{figure}

\end{document}